\documentclass[english]{article}
\usepackage[T1]{fontenc}
\usepackage[latin1]{inputenc}
\usepackage{geometry}
\usepackage{graphicx}

\makeatletter

\usepackage{babel}
\makeatother
\begin{document}

\title{\textbf{\large Arrow diagram method based on overlapping electronic
groups: \\corrections to the linked AD theorem}}

\author{Yu Wang and Lev Kantorovich$^*$}

\maketitle

\begin{center}
{\it Department of Physics, King's College London, Strand,
 London, WC2R 2LS, United Kingdom}\par

\vspace{3mm} $^*$~{\footnotesize E-mail
address:~lev.kantorovitch@kcl.ac.uk}\par
\end{center}

\begin{abstract}
Arrow diagram (AD) method (L. Kantorovich and B. Zapol, J. Chem.
Phys. \textbf{96}, 8420 (1992); \emph{ibid}, 8427) provides a
convenient means of systematic calculation of arbitrary matrix
elements, $\left\langle
\Psi\right|\widehat{O}\left|\Psi\right\rangle $, of symmetrical
operators, $\widehat{O}$, in quantum chemistry when the total
system wavefunction $\Psi$ is represented as an antisymmetrised
product of overlapping many-electron group functions, $\Phi_{A}$,
corresponding to each part (group) $A$ of the system:
$\Psi=\widehat{A}\prod_{A}\Phi_{A}$. For extended (e.g. infinite)
systems the calculation is somewhat difficult, however, as mean
values of the operators require that each term of the diagram
expansion is to be divided by the normalisation integral
$S=\left\langle \Psi\right|\left.\Psi\right\rangle $, which is
given by an AD expansion as well. A linked AD theorem suggested
previously (L. Kantorovich, Int. J. Quant. Chem. \textbf{76}, 511
(2000)) to deal with this problem is reexamined in this paper
using a simple Hartree-Fock problem of a one-dimensional ring of
infinite size which is found to be analytically solvable. We find
that corrections to the linked AD theorem are necessary in a
general case of a finite overlap between different electronic
groups. A general method of constructing these corrections in a
form of a power series expansion with respect to overlap is
suggested. It is illustrated on the ring model system.
\end{abstract}

\section{Introduction}

For a wide range of quantum-mechanical systems, the entire electronic
system can conveniently be split into a set of electron groups (EG)
such as core and valence electrons in molecules or crystals, electrons
on atoms or ions in atomic or ionic solids, core and bond electrons
in strongly covalent materials, etc. \cite{McWeeny,McWeeny-rev,EMC-1}.
Similar partitioning ideas can also be applied to separate electrons
in a cluster and environment regions to derive a particular embedding
potential for the quantum cluster \cite{EMC-1,Barandiaran-1996,Seijo-Barandiaran-1999}.
Provided that the partition scheme applied to the given system is
physically (or chemically) appropriate, one can assume that electrons
fixed to the given group spend most of their time at this group. Therefore,
to a good approximation, the wavefunction of the whole system consisting
of electronic groups $I$ can be represented as an antisymmetrised
product of wavefunctions $\Phi_{I}(X_{I})$ of every individual group:\begin{equation}
\Psi(X_{1},\cdots,X_{M})=\widehat{A}\prod_{I=1}^{M}\Phi_{I}(X_{I})\label{eq:total-psi}\end{equation}
where $M$ is the total number of EG's. Here the $I$-th group is
associated with $N_{I}$ electrons which coordinates are collected
into a set $X_{I}=(x_{1},\ldots x_{N_{I}})$. Note that the single
electron coordinates $x=(\mathbf{r},\sigma)$ include the spin coordinate
$\sigma$ as well. The antisymmetrisation operator is defined via\begin{equation}
\widehat{A}=\frac{1}{N!}\sum_{P\in S_{N}}\epsilon_{P}\widehat{P}\label{eq:operator-A}\end{equation}
where the sum runs over all $N!$ permutations $\widehat{P}$ of the
complete group of permutations $S_{N}$, the total number of electrons
in the whole system being $N=N_{1}+N_{2}+\ldots+N_{M}$ and the factor
$\epsilon_{P}=\pm1$ according to the parity of the permutation $\widehat{P}$.
It is assumed here that the group functions $\Phi_{I}(X_{I})$ are
already individually antisymmetric.

By applying expression (\ref{eq:operator-A}) for the operator
$\widehat{A}$ in Eq. (\ref{eq:total-psi}), one obtains an
expansion of the wavefunction via various functions-products. When
used in calculating matrix elements $\left\langle
\Psi\right|\widehat{O}\left|\Psi\right\rangle $ of symmetrical
(with respect to permutations of electronic coordinates) operators
$\widehat{O}\equiv\widehat{O}(x_{1},\ldots,x_{N})$ (e.g. an
external field or electron-electron interaction), a corresponding
expansion of matrix elements between various functions-products is
obtained. Earlier attempts (see, e.g.
\cite{Wilson,Klein,lyast1,lyast2,matsen1,matsen2,matsen3}) to
simplify the expansion and associate each term with a transparent
diagram have been generalised in Refs. \cite{my-AD-1,my-AD-2}
where the Arrow Diagram (AD) theory was developed. Firstly, by
exploiting a double-coset decomposition of the group $S_{N}$, many
terms in the expansion were found to be identical and the
corresponding rules to number all distinct terms were worked out
in a general case. Secondly, each distinct term in the expansion
was given by a well-defined picture (arrow diagram) which allowed
to construct simple rules to associate an analytical expression
with any of the AD's. The theory was formulated in a very general
way for arbitrary number of electron groups and explicit AD
expansions were constructed for reduced density matrices (RDM)
\cite{McWeeny,McWeeny-rev} of orders one and two (RDM-1 and
RDM-2). These enable one to calculate matrix elements of arbitrary
symmetrical one- and two-particle operators and are thus
sufficient in most cases relevant to quantum chemistry of
molecules and solids (generalisation to higher order RDM's is
cumbersome but straightforward).

In general, the success of this technique depends crucially on the
value of the overlap between different group functions $\Phi_{I}(X_{I})$
and thus their localisation in certain \emph{regions} of space (also
known as \emph{structure elements} \cite{EMC-1}). This is because
greater localisation of the group functions results in better convergence
of the AD expansion, i.e. smaller number of terms in the expansion
are to be retained. That is why the choice of electronic groups and
appropriate regions of their localisation \cite{Danyliv-LK-2004,Danyliv-LK-quartz-2004}
is so important for this method to work.

Although the AD theory formulated in \cite{my-AD-1,my-AD-2} is applicable
to any system consisting of arbitrary number of overlapping electron
groups, one can not directly use it to describe an extended (in particular,
infinite) system. Indeed, to calculate an observable \begin{equation}
\overline{O}=\frac{\left\langle \Psi\right|\widehat{O}\left|\Psi\right\rangle }{\left\langle \Psi\right|\left.\Psi\right\rangle }\label{eq:mean-value}\end{equation}
of the operator $\widehat{O}$, in the case of an extended system
one has to calculate the ratio of two AD expansions, one arising from
the matrix element $\left\langle \Psi\right|\widehat{O}\left|\Psi\right\rangle $
and another - from the normalisation integral $S=\left\langle \Psi\right|\left.\Psi\right\rangle $.
Both expansions contain very many terms; in fact, the number of terms
is infinite in the case of an infinite system (e.g. a solid).

It was argued in \cite{linked-AD} that the normalisation integral
$S$ tends to infinity with the number of electronic groups $M$.
This assertion made it possible to prove the so-called linked AD theorem
\cite{linked-AD}. It states that the observable $\overline{O}=\left\langle \Psi\right|\widehat{O}\left|\Psi\right\rangle _{c}$,
i.e. it is equal to the AD expansion for the matrix element $\left\langle \Psi\right|\widehat{O}\left|\Psi\right\rangle $
of the operator $\widehat{O}$, in which only linked (connected) ADs
are retained (subscript {}``c''), and the normalisation integral
should be dropped altogether. This expression is very attractive as
it allows one to consider a single AD expansion for arbitrary matrix
elements of operators.

It is the main objective of this paper to show that the linked AD
theorem is correct only approximately. When the overlap between different
group functions becomes larger, corrections are to be introduced.
We show this by first considering in detail a simple {}``toy'' model
of a 1D ring for which the exact analytical solution is possible.
This allows us to investigate in detail the limiting behaviour of
the AD expansion for this system when the number of groups in the
ring tends to infinity. Then, we suggest a general method of calculating
the mean values, Eq. (\ref{eq:mean-value}), in a form of corrections
to the linked AD theorem.

The paper is organised as follows. In the subsequent section we shall
briefly review the AD theory and introduce the linked AD theorem for
the normalisation integral and RDM-1. The consideration of higher
order RDM's is analogous. In section 3, we consider in detail the
toy model and obtain an exact expression for RDM-1 in the limit of
an infinite ring size. This exact results will then be compared with
that obtained using the linked AD theorem. A systematic way of constructing
corrections to the linked AD theorem is then suggested in section
4.

\section{The Arrow Diagrams theory\label{sec:The-AD-theory}}

First we consider a symmetric group $S_{N_{I}}$. Its elements permute
electronic coordinates belonging only to the $I$-th group. Joining
together all such groups we obtain a subgroup $S_{0}=S_{N_{1}}\cup S_{N_{2}}\cup\cdots\cup S_{N_{M}}$
of $S_{N}$. Any element of $S_{0}$ only interchanges electronic
coordinates within the groups, i.e. performs only intra-group permutations.
The algebraic foundation of the AD theory is based on the \emph{double
coset} (DC) decomposition of the complete symmetric group $S_{N}$
with respect to $S_{0}$, which enables to single out all inequivalent
\emph{intergroup} permutations $\widehat{P}_{qT}$ \cite{my-AD-1}:

\begin{equation}
S_{N}=\frac{1}{N_{0}}\sum_{qT}\mu_{qT}S_{0}\widehat{P}_{qT}S_{0}\label{eq:expansion-of-SN}\end{equation}
Here the sum runs over all distinct types, $q$, of operations for
intergroup permutations as well as all distinct ways, $T$, of labelling
actual groups (EG's) involved in it, and $N_{0}=N_{1}!N_{2}!\cdots\cdots N_{M}!$
is a numerical factor. $\widehat{P}_{qT}$ is a DC generator, which,
involves only inter-group permutations and, in general, can be constructed
as a product of some \emph{primitive} cycles each involving no more
than one electron from each group.

Each such a cycle is represented as a directed closed loop connecting
all groups involved in it. Thus, in general, each permutation $\widehat{P}_{qT}$
can be drawn as an arrow diagram containing a collection of closed
directed loops. These loops may pass through the given group several
times (depending on the number of electrons of the group which are
involved in the permutation $\widehat{P}_{qT}$). If the groups involved
in the diagram cannot be separated without destroying the directed
loops, the AD is called \emph{linked} or \emph{connected}. If, however,
this separation is possible than the diagram is called \emph{non-linked}
or \emph{disconnected} and it can be represented as a collection of
linked parts. The decomposition coefficients $\mu_{qT}$ can be calculated
merely by counting arrows entering and leaving each group in the AD.

By writing the antisymmetriser $\widehat{A}$ of Eq. (\ref{eq:operator-A})
using the DC generators, it is possible to obtain the diagram expansion
of the normalisation integral \cite{my-AD-2}:

\begin{equation}
S=\left\langle \widehat{A}\Phi\right|\left.\widehat{A}\Phi\right\rangle =\Lambda\sum_{qT}\epsilon_{q}\mu_{qT}\left\langle \Phi\left|\widehat{P}_{qT}\right|\Phi\right\rangle \label{eq:AD-expansion-for-S}\end{equation}
where $\Lambda=N_{0}/N!$ and $\Phi=\prod_{I}\Phi_{I}(X_{I})$ is
a product of all group functions. Eq. (\ref{eq:AD-expansion-for-S})
gives an expansion of $S$ in terms of diagrams identical to those
for the permutation group $S_{N}$ of Eq. (\ref{eq:expansion-of-SN}).
The matrix elements $\left\langle \Phi\left|\widehat{P}_{qT}\right|\Phi\right\rangle $
are represented as a product of RDM's of electronic groups involved
in the permutation $\widehat{P}_{qT}$ integrated over corresponding
electronic coordinates. If each of the group functions $\Phi_{I}$
is given as a sum of Slater determinants with molecular orbitals expanded
via some atomic orbitals (AO) basis set, then the matrix elements
$\left\langle \Phi\left|\widehat{P}_{qT}\right|\Phi\right\rangle $
are eventually expressed as a sum of products of simple overlap integrals
between the AO's of different electronic groups (see,.e.g. Ref. \cite{linked-AD}).
It is important to note here that a total contribution of a non-linked
AD is exactly equal to the product of contributions associated with
each of its linked parts \cite{my-AD-2}.

For the following, it is convenient to sort out all the terms in the
diagrammatic expansion by the number of groups involved in the AD's.
Getting rid of the common factor $\Lambda$, we can write:

\begin{equation}
\widetilde{S}=\frac{S}{\Lambda}=1+\sum_{K=2}^{M}\,\sum_{A_{1}<\cdots<A_{K}}S_{K}(A_{1},A_{2},\cdots,A_{K})\label{eq:expansion-of-S-tilda}\end{equation}
where $S_{K}(A_{1},A_{2},\cdots,A_{K})$ is the contribution of \emph{all}
AD's that correspond to intergroup permutations among $K$ different
groups with the particular labelling $A_{1},A_{2},\ldots,A_{K}$.
The unity in equation above corresponds to the trivial permutation.

For convenience, the expansion in Eq. (\ref{eq:expansion-of-S-tilda})
will be referred to as the normalisation-integral expansion. In principle,
in this expansion all groups of the entire system participate. It
is also found convenient to introduce a derivative object, $\widetilde{S}(T)$,
which is obtained from the above expansion by retaining all the AD's
which are associated only with the groups from a finite manifold $T=\{ A_{1},\ldots,A_{K}\}$
(i.e. containing groups with labels $A_{1},\ldots,A_{K}$). If $T$
comprises the whole system, we arrive at $\widetilde{S}$ of Eq. (\ref{eq:expansion-of-S-tilda}).
We shall also introduce a manifold $[T]$, i.e. an artificial system
which is obtained by {}``removing'' from the entire system all groups
comprising the set $T$. Then, $\widetilde{S}([T])$ (we shall also
use a simpler notation $\widetilde{S}[T]$ when convenient) is obtained
from Eq. (\ref{eq:expansion-of-S-tilda}) by retaining only AD's in
which any of the groups belonging to $T$ is \emph{not} present; in
other words, $\widetilde{S}([T])$ is obtained from $\widetilde{S}$
by assuming that in any of the AD's all the overlap integrals involving
groups from the manifold $T$ are equal to zero. Obviously, $\widetilde{S}$
can be considered as a particular case of $\widetilde{S}[T]$ when
the manifold $T$ is empty. We shall occasionally also use the notation
$\widetilde{S}[A_{1},\ldots,A_{K}]$ for $\widetilde{S}[T]$, where
$T=\{ A_{1},\ldots,A_{K}\}$, if we want to indicate explicitly which
particular groups are excluded.

The \emph{unnormalised} RDM-1 of the whole system,

\begin{equation}
\overline{\rho}(x;x^{\prime})=N\int\Psi(x,x_{2},\ldots,x_{N})\Psi^{*}(x^{\prime},x_{2},\ldots,x_{N})\textrm{d}x_{2}\ldots\textrm{d}x_{N}\label{eq:definition-of-RDM-1}\end{equation}
can also be written as a matrix element of a certain symmetrical one-particle
operator \cite{McWeeny,McWeeny-rev,my-AD-2}. Therefore, as in the
case of the normalisation integral, by inserting the AD expansion
for the operator $\widehat{A}$, one obtains the corresponding AD
expansion for the RDM-1 \cite{my-AD-2}. It can be constructed by
considering the AD expansion of $S$ and then modifying each diagram
by placing a small open circle, representing the variables $(x,x^{\prime})$,
in either of the three following ways: (i) on a group not involved
in the diagram; (ii) on a group involved in the diagram, and, finally,
(iii) on an arrow. Thus, each AD in the $S$ expansion serves as a
reference in building up the AD expansion for the RDM-1. Since the
contribution of any diagram, consisting of non-linked parts, is equal
to the product of contributions corresponding to each of the parts,
it can be shown \cite{linked-AD} that, in general, the AD expansion
of the RDM-1 can be represented as\begin{equation}
\widetilde{\rho}(x;x^{\prime})=\frac{\rho(x;x^{\prime})}{\Lambda}=\sum_{K=1}^{M}\,\sum_{A_{1}<\ldots<A_{K}}\,\widetilde{S}[A_{1},\ldots,A_{K}]\,\sum_{t}\rho_{K}^{t}(A_{1},\ldots,A_{K}\parallel x;x^{\prime})\label{eq:expansion-for-RDM-1}\end{equation}
 where $\rho_{K}^{t}(A_{1},\ldots,A_{K}\parallel x;x^{\prime})$ is
the sum of contributions of all AD's with an open circle which are
constructed using the particular group labelling $A_{1},A_{2},\cdots,A_{K}$.
The sum over $t$ takes account of the two positions of the open circle
on the reference AD: on an arrow and on a group. Note that each AD
with the open circle is necessarily a linked (connected) AD. One can
see from the above equation that each AD with an open circle, constructed
from groups $T=\{ A_{1},\ldots,A_{K}\}$, is multiplied by the sum
of all possible normalisation-integral AD's (i.e. AD's without the
circle), $\widetilde{S}[A_{1},\ldots,A_{K}]=\widetilde{S}[T]$, constructed
using the rest of the system $[T]$.

The RDM-1 introduced by Eq. (\ref{eq:definition-of-RDM-1}) is not
normalised to the total number of electrons in the system since the
wavefunction $\Psi$ constructed using group functions, $\Phi_{I}$,
is, in general, not normalised to unity: $S=\left\langle \Psi\right|\left.\Psi\right\rangle \neq1$.
Therefore, the true RDM-1 should be calculated according to Eq. (\ref{eq:mean-value})
as \cite{linked-AD}:\begin{equation}
\rho(x;x^{\prime})=\frac{\overline{\rho}(x;x^{\prime})}{\left\langle \Psi\right|\left.\Psi\right\rangle }=\sum_{K=1}^{M}\,\sum_{A_{1}<\ldots<A_{K}}\, f_{[T]}\,\sum_{t}\rho_{K}^{t}(A_{1},\ldots,A_{K}\parallel x;x^{\prime})\label{eq:true-RDM-1-expression}\end{equation}
where \begin{equation}
f_{[T]}=f_{[A_{1},\ldots,A_{K}]}=\frac{\widetilde{S}[A_{1},\ldots,A_{K}]}{\widetilde{S}}\label{eq:prefactors}\end{equation}
are numerical pre-factors. These depend only on the chosen set of
groups $T$. Since the normalisation integral $\widetilde{S}$ is
represented via an AD expansion, Eq. (\ref{eq:AD-expansion-for-S}),
one can see that the pre-factors are given as a ratio of two AD expansions,
each containing a very large (infinite) number of terms for a large
(infinite) system. Thus, it follows from the last passage in Eq. (\ref{eq:true-RDM-1-expression}),
that the true RDM-1 is represented as a sum of all linked AD's with
the open circle, $\rho_{K}^{t}(A_{1},\ldots,A_{K}\parallel x;x^{\prime})$,
multiplied by numerical pre-factors, $f_{[A_{1},\ldots,A_{K}]}$.
The calculation of the pre-factors poses the main problem in applying
the AD theory to infinite or even large system (for small systems
all AD's can be accounted for explicitly and thus $\rho$ can easily
be calculated, at least in principle).

It was argued in \cite{linked-AD} that in the limit of an infinite
system the pre-factors $f_{[T]}$ tend exactly to unity for any choice
of the groups $T=\{ A_{1},\ldots,A_{K}\}$. As a result, it was argued
that the true RDM-1 can be represented as a single AD expansion containing
only AD's with the open circle which, as was mentioned earlier, are
all linked AD's. The main argument put forward to prove this so-called
'linked-AD theorem' was that any $\widetilde{S}[A_{1},\ldots,A_{K}]$
(and, in particular, $\widetilde{S}$) tends to infinity when the
number of electronic groups $M$ tends to infinity.

The main objective of this paper is to show that the situation is
more subtle. Although the argument about the limiting behaviour of
$\widetilde{S}[T]$ put forward in \cite{linked-AD} may seem quite
plausible, in some cases, as will be shown in the next section by
considering an exactly solvable 'toy' model, $\widetilde{S}[T]$ may
have a zero limit. The main reason for this is that contributions
of different AD's in the expansion have alternating signs, the point
overlooked previously. As a result, the pre-factors $f_{[T]}$ do
not necessarily tend to unity, but may take a different value depending
on the given system.

\section{1D toy model: a Hartree-Fock ring}

\begin{figure}
\begin{center}\includegraphics[%
  height=6cm]{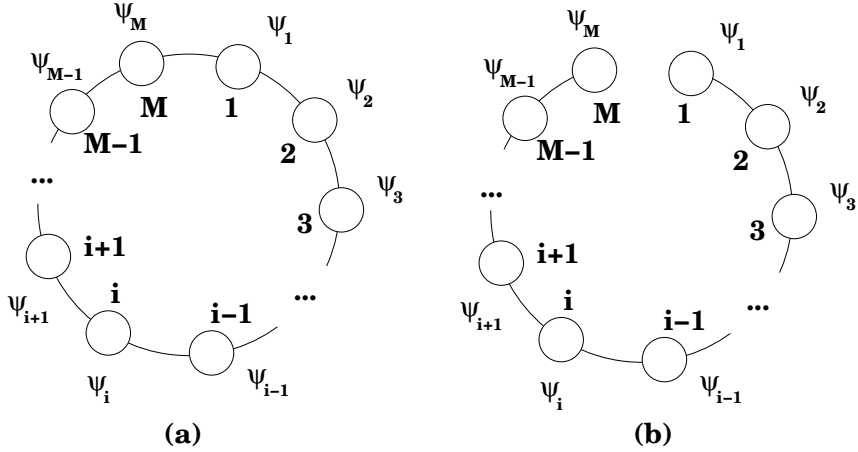}\end{center}

\caption{The 1D 'toy' model: (\emph{a}) a HF ring consisting of equidistant
one-electron orbitals $\psi_{1}$, $\psi_{2}$, $\ldots$, $\psi_{M}$
and (\emph{b}) the corresponding chain model in which there is no
overlap between the first and the last orbitals in the ring. \label{cap:Our-'toy'-model}}
\end{figure}

Let us consider a ring of $M$ equally spaced one-electron groups
as shown in Fig. \ref{cap:Our-'toy'-model} (\emph{a}). %
 Each group is described by a single real $s$-type normalised wavefunction
$\psi_{i}(x)$ ($i=1,2,\ldots,M$) localised around the group centre.
All groups are identical, i.e. every localised function $\psi_{i}(x)$
can be obtained by a corresponding translation of the neighbouring
function. Note that $N=M$ for this model and the ring wavefunction
of Eq. (\ref{eq:total-psi}) represents essentially a single Slater
determinant $\Psi\propto\det\left\{ \psi_{1}(x_{1})\cdots\psi_{M}(x_{M})\right\} $.
Hence, this theory should be equivalent to the Hartree-Fock (HF) method.

We assume that only neighbouring functions overlap. Due to the symmetry
of the model, only one parameter, namely the overlap integral $\sigma=\int\psi_{i}(x)\psi_{i+1}(x)\textrm{d}x$
between any neighbouring groups, is required to characterise the system;
the actual spatial form of $\psi_{i}(x)$ is not actually needed.

We shall show that the RDM-1 of this system can be calculated analytically
for any $M$. Most remarkably, it will be shown that the $M\rightarrow\infty$
limit is also analytically attainable. Using this result, we will
analyse the linked-AD theorem. We shall find that some corrections
are required.

Two independent methods will be used to consider the toy model.

\subsection{Finite ring: method based on the inverse of the overlap matrix}

It is first instructive to perform the calculation using a method
which is completely independent of the AD theory. This is indeed possible
since an expression for the RDM-1 of any HF system can be written
explicitly via orbitals and their overlap matrix \cite{McWeeny}.
Taking into account that only overlap between nearest neighbours exists,
we obtain for the ring electron density (the diagonal elements of
the RDM-1):\[
\rho(x)=\sum_{i,j=1}^{N}\psi_{i}(x)(\mathbf{S}^{-1})_{ij}\psi_{j}(x)\]

\begin{equation}
=\sum_{i=1}^{N}\left[\psi_{i}(x)(\mathbf{S}^{-1})_{ii}\psi_{i}(x)+2\psi_{i}(x)(\mathbf{S}^{-1})_{i,i+1}\psi_{i+1}(x)\right]\label{eq:ro-via-inverse-S}\end{equation}
Here $\mathbf{S}^{-1}$ is the inverse of the overlap matrix

\begin{equation}
\mathbf{S}=\left[\begin{array}{ccccccc}
1 & \sigma &  &  &  &  & \sigma\\
\sigma & 1 & \sigma\\
 & \ddots & \ddots & \ddots\\
 &  & \ddots & \ddots & \ddots\\
 &  &  & \ddots & \ddots & \ddots\\
 &  &  &  & \sigma & 1 & \sigma\\
\sigma &  &  &  &  & \sigma & 1\end{array}\right]_{N\times N}\label{eq:overlap-matrix}\end{equation}
between the ring orbitals. Note that all the omitted elements in the
above expression are zeros. This convention is adopted hereafter.
Then, the normalisation integral:

\begin{equation}
G_{N}^{Ring}=\textrm{det}\mathbf{S}\label{eq:G-ring}\end{equation}
is simply given by the determinant of the matrix $\mathbf{S}$ of
Eq. (\ref{eq:overlap-matrix}). To avoid a possible confusion with
the overlap matrix, we use for the normalisation integral a different
notation here, $G_{N}^{Ring}$.

For the convenience of the following calculation, we also introduce
a similar \emph{chain} system shown in Fig. \ref{cap:Our-'toy'-model}
(\emph{b}) in which the overlap between the groups $1$ and $M$ is
equal to zero. The normalisation integral for the chain is

\begin{equation}
G_{N}^{Chain}=\left|\begin{array}{ccccccc}
1 & \sigma\\
\sigma & 1 & \sigma\\
 & \ddots & \ddots & \ddots\\
 &  & \ddots & \ddots & \ddots\\
 &  &  & \ddots & \ddots & \ddots\\
 &  &  &  & \sigma & 1 & \sigma\\
 &  &  &  &  & \sigma & 1\end{array}\right|_{N\times N}\label{eq:G-chain}\end{equation}

Owing to the symmetry of the above introduced determinants, it is
possible to derive the following recurrence relations:\begin{equation}
G_{N}^{Ring}=G_{N-1}^{Chain}-2\sigma^{2}G_{N-2}^{Chain}+2\left(-1\right)^{N+1}\sigma^{N}\label{eq:recurrence-1}\end{equation}

\begin{equation}
G_{N}^{Chain}=G_{N-1}^{Chain}-\sigma^{2}G_{N-2}^{Chain}\label{eq:recurrence-2}\end{equation}
The first one is obtained by opening the determinant of the matrix
(\ref{eq:overlap-matrix}) along its first row (or column), while
the second relation is obtained in the same way from Eq. (\ref{eq:G-chain}).
Combined with the obvious 'initial' expressions $G_{1}^{Ring}=G_{1}^{Chain}=1$
and $G_{2}^{Chain}=1-\sigma^{2}$, the above relations allow one to
calculate the $G_{N}^{Chain}$ and $G_{N}^{Ring}$ for any value of
$N$. For instance, we obtain, \begin{equation}
G_{6}^{Ring}=1-6\sigma^{2}+9\sigma^{4}-4\sigma^{6}\label{eq:G-ring-6}\end{equation}
\begin{equation}
G_{7}^{Ring}=1-7\sigma^{2}+14\sigma^{4}-7\sigma^{6}+2\sigma^{7}\label{eq:G-ring-7}\end{equation}
\begin{equation}
G_{6}^{Chain}=1-5\sigma^{2}+6\sigma^{4}-\sigma^{6}\label{eq:G-chain-6}\end{equation}
\begin{equation}
G_{7}^{Chain}=1-6\sigma^{2}+10\sigma^{4}-4\sigma^{6}\label{eq:G-chain-7}\end{equation}
Note the alternating sign in every next term in the expressions above.
Also, in agreement with \cite{linked-AD}, one can see that numerical
pre-factors in each term grow with increase of $N$.

It is now straightforward to investigate the limiting behaviour of
these quantities when the number of groups tends to infinity (the
$N\rightarrow\infty$ limit). We find, using a simple numerical calculation,
that both of them tend to \emph{zero} with increase of $N$ for \emph{any}
value of $\sigma<0.5$ (it will be clear later on that the limit does
not exist for $\sigma\geq0.5$). Since both $G_{N}^{Chain}$ and $G_{N}^{Ring}$
correspond to the normalisation integrals for the two systems, one
can see that this limiting behaviour is quite different from what
should have been expected from the analysis performed in Ref. \cite{linked-AD}.
This is explained by the fact that both $\lim_{N\rightarrow\infty}G_{N}^{Ring}$
and $\lim_{N\rightarrow\infty}G_{N}^{Chain}$ are given by infinite
series with alternating terms. The limiting behaviour of each of the
series is not straightforward in spite of the fact that the pre-factors
to \emph{each} term $\sigma^{n}$($n=2,3,\ldots$) tend to infinity
when $N\rightarrow\infty$.

In order to calculate the RDM-1 (\ref{eq:ro-via-inverse-S}), one
also need the corresponding elements of the inverse of the overlap
matrix. Using the explicit structure of the matrix $\mathbf{S}$ given
in Eq. (\ref{eq:overlap-matrix}), one obtains:

\begin{equation}
\left(\mathbf{S}^{-1}\right)_{i,i}=\frac{C_{i,i}}{\textrm{det}\mathbf{S}}=\frac{G_{N-1}^{Chain}}{G_{N}^{Ring}}\label{eq:S(-1)-ii}\end{equation}

\begin{equation}
\left(\mathbf{S}^{-1}\right)_{i,i+1}=\left(\mathbf{S}^{-1}\right)_{i+1,i}=\frac{C_{i,i+1}}{\textrm{det}\mathbf{S}}=\frac{-\sigma G_{N-2}^{Chain}+\left(-1\right)^{N+1}\sigma^{N-1}}{G_{N}^{Ring}}\label{eq:S(-1)-ij}\end{equation}
where $C_{i,i}$ and $C_{i,i+1}$ are cofactors of $\textrm{S}_{i,i}$
and $\textrm{S}_{i+1,i}$ respectively.

Using Eqs. (\ref{eq:S(-1)-ii}) and (\ref{eq:S(-1)-ij}), we obtain
for the electron density (\ref{eq:ro-via-inverse-S}):

\[
\rho(x)=\left(\frac{G_{N-1}^{Chain}}{G_{N}^{Ring}}\right)\sum_{i=1}^{N}\psi_{i}^{2}(x)\]

\begin{equation}
-2\sigma\left(\frac{G_{N-2}^{Chain}}{G_{N}^{Ring}}\right)\sum_{i=1}^{N}\psi_{i}(x)\psi_{i+1}(x)-\frac{2}{\sigma}\left(\frac{\left(-\sigma\right)^{N}}{G_{N}^{Ring}}\right)\sum_{i=1}^{N}\psi_{i}(x)\psi_{i+1}(x)\label{eq:electron-density-ring}\end{equation}

The above expression allows one to calculate the electron density
of the ring with arbitrary number of groups. The $N\rightarrow\infty$
limit will be considered separately in section \ref{sub:Infinite-ring}.

\subsection{Finite ring: method based on the AD theory}

We shall show in this section that the same expressions as derived
in the previous subsection can also be obtained using a much simpler
algebra of the AD theory.

\begin{figure}
\begin{center}\includegraphics[%
  height=3cm]{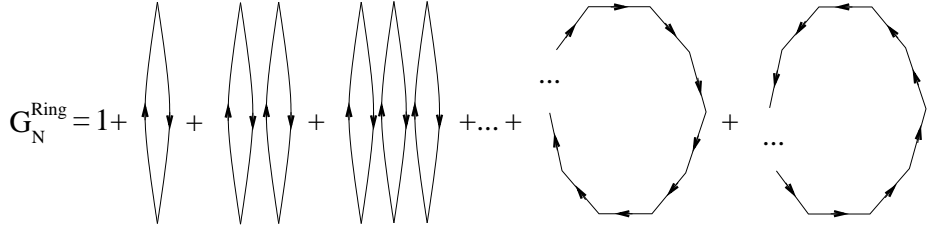}\end{center}

\caption{The AD expansion for the ring normalisation integral, $\widetilde{S}$.
The ring consists of a finite number $N$ of one-electron groups.
\label{cap:S-expansion-for-ring}}
\end{figure}

To this end, we first consider the normalisation integral $\widetilde{S}\equiv G_{N}^{Ring}$
of the ring. Since all groups contain a single electron and only nearest
groups have non-zero overlap, the AD expansion contains essentially
only bubble-like AD's between two adjacent groups (including all non-linked
AD's constructed out of them) and two AD's connecting \emph{all} groups
(a $N$-vertex polygon) with opposite direction of arrows, as depicted
in Fig. \ref{cap:S-expansion-for-ring}. An AD expansion for the $G_{N}^{Chain}$
looks similarly. There are two differences: (i) bubble AD's containing
the groups $1$ and $N$ are missing, and (ii) it does not contain
the two polygon diagrams. Recall that there is no {}``connection''
between the groups 1 and $N$ in the chain. %

\begin{figure}
\begin{center}\includegraphics[%
  height=3cm]{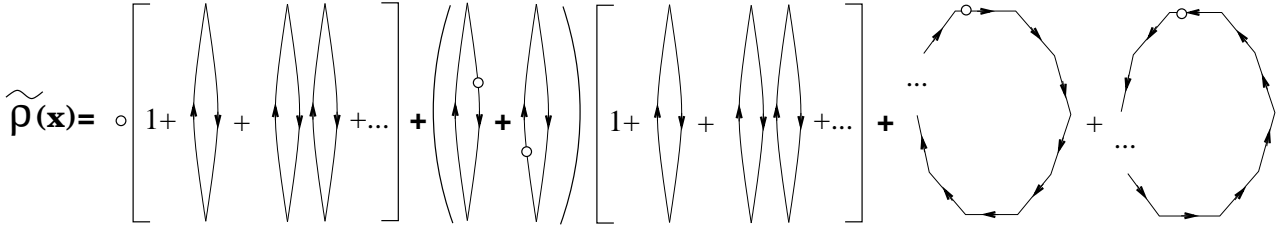}\end{center}

\caption{The AD expansion for the diagonal elements of the ring RDM-1, $\widetilde{\rho}(x)\equiv\widetilde{\rho}(x;x)$,
Eq. (\ref{eq:expansion-for-RDM-1}). \label{cap:RDM-1-for-ring}}
\end{figure}

An AD expansion for the electron density can also be written explicitly
as shown in Fig. \ref{cap:RDM-1-for-ring}. It contains three terms.
 The first term is associated with an open circle AD, associated with,
say, group $i$, multiplied by all bubble AD's (shown in the square
brackets) which are constructed out of the rest $N-1$ groups $1,\ldots,i-1,i+1,\ldots,N$.
The contribution of the open circle $\rho_{i}^{(\circ)}(x)=\psi_{i}^{2}(x)$,
while the bubble AD's in the square brackets all amount to $G_{N-1}^{Chain}$
(and are the same for each $i$), since they represent all AD's for
a chain with $N-1$ groups. Finally, one has to sum over all values
of $i$ (all positions of the open circle). Hence, the first term
contributes $G_{N-1}^{Chain}\sum_{i=1}^{N}\rho_{i}^{(\circ)}(x)$
to the density $\widetilde{\rho}(x)$.

The second term in Fig. \ref{cap:RDM-1-for-ring} contains two bubble
diagrams with an open circle on both of its arrows, multiplied by
all possible bubble AD's constructed out of the rest $N-2$ groups.
Assuming that the bubble AD is taken between the groups $i$ and $i+1$,
we obtain the contribution $\rho_{i,i+1}^{(bubble)}(x)=-\sigma\psi_{i}(x)\psi_{i+1}(x)$
for it. Since the contribution of all groups $1,\ldots,i-1,i+2,\ldots,N$
in the square brackets in the second term is equal to $G_{N-2}^{Chain}$
(which is the same for any pair $i$, $i+1$), the final contribution
of the second term in Fig. \ref{cap:RDM-1-for-ring} becomes $2G_{N-2}^{Chain}\sum_{i=1}^{N}\rho_{i,i+1}^{(bubble)}(x)$,
where we have also summed over all pairs, i.e. all possible positions
of the open circle on the bubble AD's (recall, that the bubble AD's
can be taken only between nearest neighbours in our model).

Finally, the last two terms in Fig. \ref{cap:RDM-1-for-ring} are
represented by polygons with an open circle. Each such a diagram contributes
$\rho_{i,i+1}^{(poly)}=\left(-\sigma\right)^{N-1}\psi_{i}(x)\psi_{i+1}(x)$
if the open circle is positioned on the arrow connecting groups $i$
and $i+1$. The total contribution of the polygons is obtained by
summing over all possible positions of the circle and multiplying
by a factor of two since each AD accepts two directions of the arrows.

Summing all three contributions, we arrive at the following expression
for the unnormalised electron density: \begin{equation}
\widetilde{\rho}(x)=G_{N-1}^{Chain}\sum_{i=1}^{N}\rho_{i}^{(\circ)}(x)+2G_{N-2}^{Chain}\sum_{i=1}^{N}\rho_{i,i+1}^{(bubble)}(x)+2\sum_{i=1}^{N}\rho_{i,i+1}^{(poly)}(x)\label{eq:AD-ring}\end{equation}
Dividing this expression by $G_{N}^{Ring}$, one obtains the normalised
electron density $\rho(x)$, which, as can easily be seen, appears
to be identical to that of Eq. (\ref{eq:electron-density-ring}) obtained
in the previous section using Slater determinants.

To finish the proof of the equivalence of the two methods, we only
need to re-derive the recurrence relations (\ref{eq:recurrence-1})
and (\ref{eq:recurrence-2}) using exclusively the AD theory.

Consider first the expansion for $G_{N}^{Chain}$ which contains only
zero, one, two, etc. bubble diagrams (see the bubble-containing terms
in Fig. \ref{cap:S-expansion-for-ring}). Recall that the chain is
broken between groups 1 and $N$. Let us {}``fix'' group 1. Then,
$G_{N}^{Chain}$ can be written as the sum of two contributions: (i)
due to all diagrams involving group 1 and (ii) due to the rest of
them which do not involve it. The first contribution is simply a single
bubble AD between groups 1 and 2, equal to $-\sigma^{2}$, times all
the bubble AD's made of all the other groups, i.e. the whole contribution
is $\left(-\sigma^{2}\right)G_{N-2}^{Chain}$. The contribution of
all AD's which do not involve group $1$ is simply equal to $G_{N-1}^{Chain}$.
One can immediately recognise the recurrence relation (\ref{eq:recurrence-2}).

Eq. (\ref{eq:recurrence-1}) is proven in a similar way. Again, we
first {}``fix'' group 1. Then, the sum of all AD's shown in Fig.
\ref{cap:S-expansion-for-ring} is equal to the sum of three contributions:
(i) $G_{N-1}^{Chain}$ due to all AD's which do not involve group
$1$; (ii) all the bubble AD's involving it, given by $2\left(-\sigma^{2}\right)G_{N-2}^{Chain}$
analogously to the case of the chain discussed above (the factor of
two arises due to the fact that one can construct two bubble AD's
with group $1$, namely those involving group pairs $1,N$ and $1,2$);
(iii) the contribution $-2\left(-\sigma\right)^{N}$of the two polygon
AD's. Summing all three terms, one arrives exactly at Eq. (\ref{eq:recurrence-1}),
as required.

\subsection{Infinite ring\label{sub:Infinite-ring}}

Our task now is to calculate the pre-factors (in the round brackets)
in the expression for the density of Eq. (\ref{eq:electron-density-ring}).
To do this, it was found convenient to introduce the following quantity:
\begin{equation}
P_{N}=\sigma^{2}\left(\frac{G_{N-1}^{Chain}}{G_{N}^{Chain}}\right)\label{eq:Pn-def}\end{equation}
Using explicit expressions for the $G_{2}^{Chain}$ and $G_{1}^{Chain}$,
we get $P_{2}=\sigma^{2}/\left(1-\sigma^{2}\right)$. Using the recurrence
relation (\ref{eq:recurrence-2}) for $G_{N}^{Chain}$, a very simple
recurrence relation for $P_{N}$ can also be derived:

\begin{equation}
P_{N}=\frac{\sigma^{2}}{1-P_{N-1}}\label{eq:recurr-for-Pn}\end{equation}

The introduced quantity is very useful in two respects. Firstly, the
first two pre-factors in Eq. (\ref{eq:electron-density-ring}) can
be directly expressed via it:

\begin{equation}
\frac{G_{N-1}^{Chain}}{G_{N}^{Ring}}=\frac{1}{1-2P_{N-1}+2\sigma\chi_{N-1}}\label{eq:ratio-1}\end{equation}

\begin{equation}
\frac{G_{N-2}^{Chain}}{G_{N}^{Ring}}=\frac{\sigma^{-2}P_{N-1}}{1-2P_{N-1}+2\sigma\chi_{N-1}}\label{eq:ratio-2}\end{equation}
where \begin{equation}
\chi_{N}=\frac{(-\sigma)^{N}}{G_{N}^{Chain}}\label{eq:ratio-3}\end{equation}
Secondly, it has a finite $N\rightarrow\infty$ limit. Indeed, assuming
that such a limit exists, we obtain from Eq. (\ref{eq:recurr-for-Pn})
a simple quadratic equation $P_{\infty}(1-P_{\infty})=\sigma^{2}$,
where $P_{\infty}=\lim_{N\rightarrow\infty}P_{N}$. The roots of this
equation are $\frac{1}{2}\left(1\pm\sqrt{1-4\sigma^{2}}\right)$.
To choose the correct sign, we consider the case of $\sigma=0$, when
$G_{N}^{Chain}=1$ for any $N$, and thus $P_{N}=0$. Only the root
with the minus sign, \begin{equation}
P_{\infty}=\frac{1-\sqrt{1-4\sigma^{2}}}{2}\label{eq:P-infinity}\end{equation}
 satisfies this condition. It also appears that a definite real limit
exists only for $\sigma<\frac{1}{2}$ which corresponds to a not very
large overlap. This restriction is a consequence of the nearest-neighbour
approximation adopted in our toy model. Note in passing that $P_{N}>0$
for any $N$.

Finally, to finish the calculation, we have to consider the limiting
behaviour of the ratios $\chi_{N}$ and \begin{equation}
\xi_{N}=\frac{(-\sigma)^{N}}{G_{N}^{Ring}}=\frac{-\sigma\chi_{N-1}}{1-2P_{N-1}+2\sigma\chi_{N-1}}\label{eq:ratio-4}\end{equation}
which enter Eqs. (\ref{eq:ratio-1}), (\ref{eq:ratio-2}) and (\ref{eq:electron-density-ring}),
respectively. To calculate the $\lim_{N\rightarrow\infty}\chi_{N}$,
we use the definition (\ref{eq:Pn-def}) of $P_{N}$ to write:

\[
G_{N-1}^{Chain}=\frac{\left(\sigma^{2}\right)^{N-3}}{P_{N-1}P_{N-2}\cdots P_{3}}\cdot G_{2}^{Chain}>\left(\frac{\sigma^{2}}{P_{\infty}}\right)^{N-3}\cdot G_{2}^{Chain}\]
since $P_{N}<P_{\infty}$ for any finite value of $N$ and any $0\leq\sigma<\frac{1}{2}$.
Since $G_{2}^{Chain}=1-\sigma^{2}>0$ and any of the quantities $P_{K}$
are positive, one can write

\[
0<\frac{\sigma^{N}}{G_{N-1}^{Chain}}<\frac{\sigma^{3}}{G_{2}^{Chain}}\cdot\left(\frac{P_{\infty}}{\sigma}\right)^{N-3}\]
Because $\frac{P_{\infty}}{\sigma}<1$ for any $0\leq\sigma<\frac{1}{2}$,
we obtain $\lim_{N\rightarrow\infty}\left(\frac{P_{\infty}}{\sigma}\right)^{N-3}=0$
and then

\begin{equation}
\lim_{N\rightarrow\infty}\left(\frac{\sigma^{N}}{G_{N-1}^{Chain}}\right)=0\label{eq:first-limit}\end{equation}
Hence, $\chi_{N}$ of Eq. (\ref{eq:ratio-3}) has a zero limit and
disappears in the ratios (\ref{eq:ratio-1}) and (\ref{eq:ratio-2}).
Consequently, $\xi_{N}\rightarrow0$ too (see Eq. (\ref{eq:ratio-4})),
so that the third term in the right hand side of Eq. (\ref{eq:electron-density-ring})
also does not contribute when the limit of an infinite ring is taken.

Finally, combining all contributions, we obtain the following \emph{exact}
result for the infinite ring:\begin{equation}
\rho(x)=\sum_{i=1}^{N}f_{[i]}\rho_{i}^{(\circ)}(x)+2\sum_{i=1}^{N}f_{[i,i+1]}\rho_{i,i+1}^{(bubble)}(x)\label{eq:RDM-infinite-ring-1}\end{equation}
 where the corresponding expressions for the two pre-factors are:
\begin{equation}
f_{[i]}=\frac{1}{\sqrt{1-4\sigma^{2}}}\label{eq:prefactor-1}\end{equation}
\begin{equation}
f_{[i,i+1]}=\frac{1}{2\sigma^{2}}\left(\frac{1}{\sqrt{1-4\sigma^{2}}}-1\right)\label{eq:prefactor-2}\end{equation}

\begin{figure}
\begin{center}\includegraphics[%
  height=7cm]{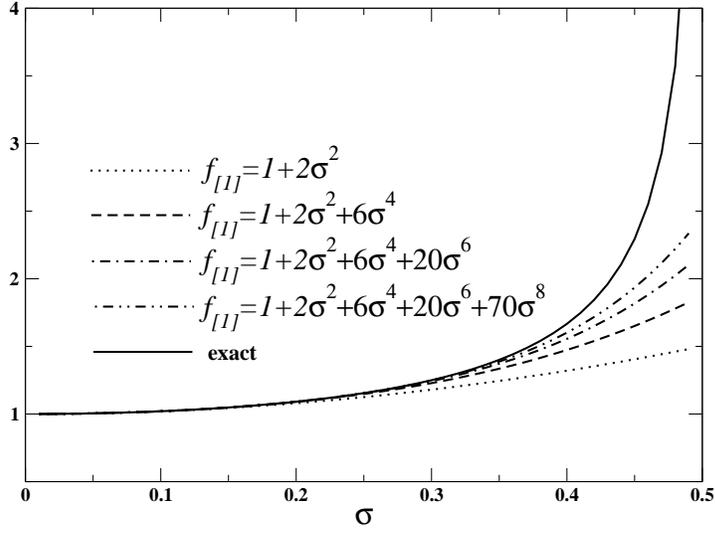}\end{center}

\caption{Comparison of the exact expression for the pre-factor $f_{[1]}(\sigma)$,
Eq. (\ref{eq:prefactor-1}), with its series expansion, Eq. (\ref{eq:expansion_for_f1}).\label{cap:f1_exact_versus_expansion}}
\end{figure}

\begin{figure}
\begin{center}\includegraphics[%
  height=7cm]{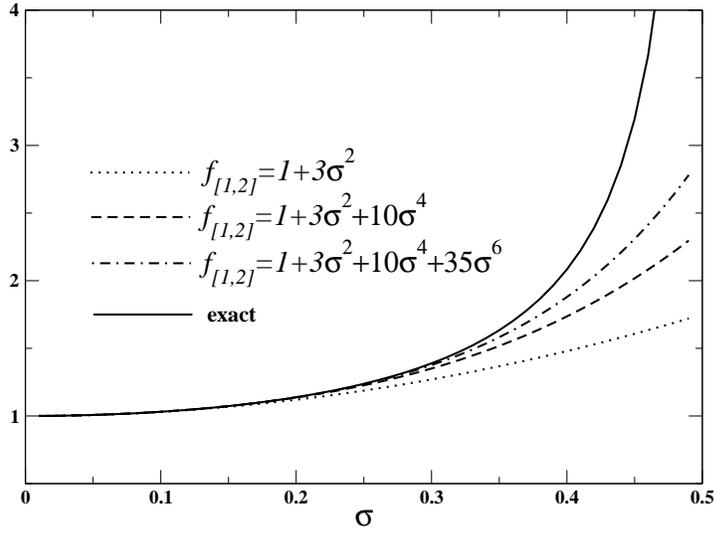}\end{center}

\caption{Comparison of the exact expression for the pre-factor $f_{[1,2]}(\sigma)$,
Eq. (\ref{eq:prefactor-2}), with its series expansion, Eq. (\ref{eq:expansion_for_f12}).
\label{cap:f12_exact_versus_expansion}}
\end{figure}

The above obtained expression for the normalised density can be compared
directly with the general formula (\ref{eq:true-RDM-1-expression}).
According to the linked-AD theorem \cite{linked-AD}, both pre-factors
are equal to unity for any $\sigma$. We see, however, that is not
the case in general. The actual pre-factors $f_{[i]}$ and $f_{[i,i+1]}$
are plotted in Figs. \ref{cap:f1_exact_versus_expansion} and \ref{cap:f12_exact_versus_expansion},
respectively, for the whole range of the overlap integral $\sigma$
by solid lines. %

One can clearly see that for $0\leq\sigma\preceq0.3$, i.e. for the
small and intermediate overlap, both pre-factors are approximately
equal to unity, which is the value predicted by the linked-AD theorem.
However, when the overlap between neighbouring group functions is
larger, $0.3\prec\sigma<0.5$, both pre-factors shoot up to infinity.
Note that the singularity at $\sigma=0.5$ is somewhat artificial
and is removed if the nearest-neighbour approximation for the overlap
is lifted.

Thus, we conclude, that the linked-AD theorem works well in the region
of not very large overlap, i.e. when the group functions are sufficiently
localised. If the overlap is large, then corrections to the linked-AD
theorem are necessary. A general method for building up such corrections
will be suggested in the next section.

\section{Corrections to the linked-AD theorem}

The idea of the method to be proposed in this section is based on
the power series expansion for the pre-factors. In order to illustrate
the general method, it is instructive to consider first a simple example
of the familiar 1D toy model.

\subsection{Series expansion for the 1D toy model\label{sub:Series-expansion-for-toy-model}}

Let us consider the 1D toy model (an infinite ring) for which the
exact solution is known. Fix group 1. Then, the normalisation integral
for the whole ring, $\widetilde{S}$, can be constructed as a sum
of three terms: (i) all AD's of the rest of the system, $\widetilde{S}[1]$,
where $T=[1]$ corresponds to the ring with group 1 {}``removed'',
i.e. this group does not have any overlap with its neighbours; (ii)
a bubble AD between groups 1 and 2 (contributing $-\sigma^{2}$) multiplied
by all possible bubble AD's due to all other groups, $\widetilde{S}[1,2]$
(as usual, $[1,2]$ denotes the system in which groups 1 and 2 are
removed), and similarly (iii) a bubble AD between group 1 and the
last group in the ring (the other neighbour of group 1) times all
the bubble AD's of the rest of the system. The last two contributions
are identical. Note that we do not consider the polygon here as we
know (section \ref{sub:Infinite-ring}) that in the limit of an infinite
system its contribution vanishes. Thus, one can write: $\widetilde{S}=\widetilde{S}[1]-2\sigma^{2}\widetilde{S}[1,2]$.
Dividing both sides of this equation by $\widetilde{S}$, we obtain
an equation connecting pre-factors: \begin{equation}
1=f_{[1]}-2\sigma^{2}f_{[1,2]}\label{eq:for_S}\end{equation}
 Repeating the above procedure for the system $[1]$ and fixing group
2, we similarly obtain: $\widetilde{S}[1]=\widetilde{S}[1,2]-\sigma^{2}\widetilde{S}[1,2,3]$,
where $[1,2,3]$ denotes a system in which groups 1, 2 and 3 are removed.
In fact, the above equation also follows from the recurrence relation
(\ref{eq:recurrence-2}). Dividing the last equation by $\widetilde{S}$,
we obtain another relationship between pre-factors: \begin{equation}
f_{[1]}=f_{[1,2]}-\sigma^{2}f_{[1,2,3]}\label{eq:for_f1}\end{equation}
Continuing this procedure, one can write an infinite series of relationships
in which every time a new pre-factor appears, for instance: \begin{equation}
f_{[1,2]}=f_{[1,2,3]}-\sigma^{2}f_{[1,2,3,4]}\label{eq:for_f12}\end{equation}

\begin{equation}
f_{[1,2,3]}=f_{[1,2,3,4]}-\sigma^{2}f_{[1,2,3,4,5]}\label{eq:for_f123}\end{equation}
Next, we assume that the pre-factors are well-defined functions of
the overlap, $\sigma$. Hence, they all can be expanded in a power
series:\begin{equation}
f_{[T]}=1+\sum_{n=1}^{\infty}a_{n}^{[T]}\sigma^{n}\label{eq:expansion_for_f}\end{equation}
where $T$ is either $1$ or $1,2$ or $1,2,3$ and so on. Note that
the expansion starts from the unity for any $[T]$ since $f_{[T]}=1$
for zero overlap ($\sigma=0$). Substituting these expansions into
Eqs. (\ref{eq:for_S})-(\ref{eq:for_f123}), one can recursively recover
the expansion coefficients up to a certain order by comparing terms
of the same power of $\sigma$. Indeed, it follows from Eq. (\ref{eq:for_S})
that $a_{1}^{[1]}=0$, $a_{2}^{[1]}=2$ and $a_{n}^{[1]}=2a_{n-2}^{[1,2]}$
for any $n\geq3$. From Eq. (\ref{eq:for_f1}) we get: $a_{1}^{[1,2]}=a_{1}^{[1]}=0$,
$a_{2}^{[1,2]}=a_{2}^{[1]}+1=3$ and $a_{n}^{[1]}-a_{n}^{[1,2]}+a_{n-2}^{[1,2,3]}=0$
for any $n\geq3$. Similarly, we obtain from Eq. (\ref{eq:for_f12})
that: $a_{1}^{[1,2,3]}=a_{1}^{[1,2]}=0$, $a_{2}^{[1,2,3]}=a_{2}^{[1,2]}+1=4$
and $a_{n}^{[1,2]}-a_{n}^{[1,2,3]}+a_{n-2}^{[1,2,3,4]}=0$ for $n\geq3$.
Combining the above recursive relationships, new coefficients can
be obtained, e.g. $a_{3}^{[1,2]}=0$, $a_{4}^{[1,2]}=10$, $a_{3}^{[1]}=0$,
$a_{4}^{[1]}=6$, $a_{5}^{[1]}=0$, $a_{6}^{[1]}=20$. Using Eq. (\ref{eq:for_f123})
enables calculation of more coefficients. This way we obtain the first
several terms in the expansion of the two pre-factors that are required
for the 1D toy model: \begin{equation}
f_{[1]}=1+2\sigma^{2}+6\sigma^{4}+20\sigma^{6}+70\sigma^{8}+\ldots\label{eq:expansion_for_f1}\end{equation}

\begin{equation}
f_{[1,2]}=1+3\sigma^{2}+10\sigma^{4}+35\sigma^{6}+\ldots\label{eq:expansion_for_f12}\end{equation}
It can easily be seen by expanding in a power series the exact expressions
(\ref{eq:prefactor-1}) and (\ref{eq:prefactor-2}) for the pre-factors,
that the above expansions are indeed correct. To obtain more terms
in the expansion by using this method, one has to consider higher
order pre-factors. The expansions for $f_{[1]}$ and $f_{[1,2]}$
obtained above are compared with the exact expressions of Eqs. (\ref{eq:prefactor-1})
and (\ref{eq:prefactor-2}), respectively, in Figs. \ref{cap:f1_exact_versus_expansion}
and \ref{cap:f12_exact_versus_expansion}. One can see that in both
cases reasonable approximations to the exact values of the pre-factors
are obtained using the power series expansion in the cases of small
and intermediate overlap. It is also apparent that one has to go to
higher orders in the expansions when the overlap is much larger. In
particular, the power series expansions do not show sharp increase
of the pre-factors closer to the critical overlap $\sigma=\frac{1}{2}$.

\subsection{General method\label{sub:General-method}}

\begin{figure}
\begin{center}\includegraphics[%
  height=6cm]{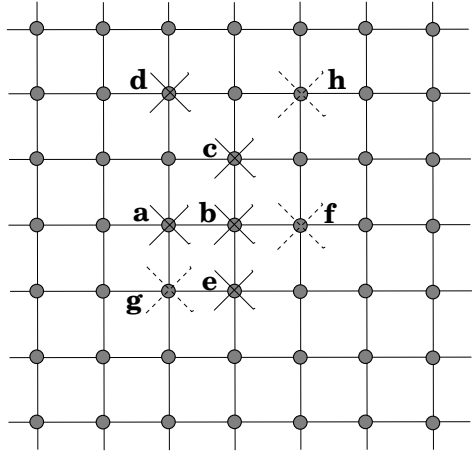}\end{center}

\caption{A schematic of an infinite 2D lattice of electronic groups. The solid
crosses indicate the {}``removed'' groups that comprise the set
$T=\{ a,b,c,d,e\}$, whereas dashed crosses indicate additional groups
to be {}``removed'' which form a set $\Delta T=\{ f,g,h\}$. \label{cap:An-infinite-2D}}
\end{figure}

The idea of the series expansion method of the previous subsection
can actually be generalised for an arbitrary system containing groups
with any numbers of electrons, i.e. for the most general wavefunction
of Eq. (\ref{eq:total-psi}). Indeed, what is needed is a way to connect
the pre-factor $f_{[T]}=\widetilde{S}[T]/\widetilde{S}$ associated
with some set $T=\left\{ A_{1},A_{2},\ldots,A_{K}\right\} $ of groups
(not necessarily nearest neighbours) with the pre-factor $f_{\left[T^{\prime}\right]}=\widetilde{S}[T^{\prime}]/\widetilde{S}$
associated with a \emph{smaller} set $[T^{\prime}]=[T\bigcup\Delta T]=[T+\Delta T]$.
The set $T^{\prime}$ contains $L$ additional groups forming a set
$\Delta T=\left\{ B_{1},B_{2},\ldots,B_{L}\right\} $, see an illustration
in Fig. \ref{cap:An-infinite-2D}.

\begin{figure}
\begin{center}\includegraphics[%
  height=6cm]{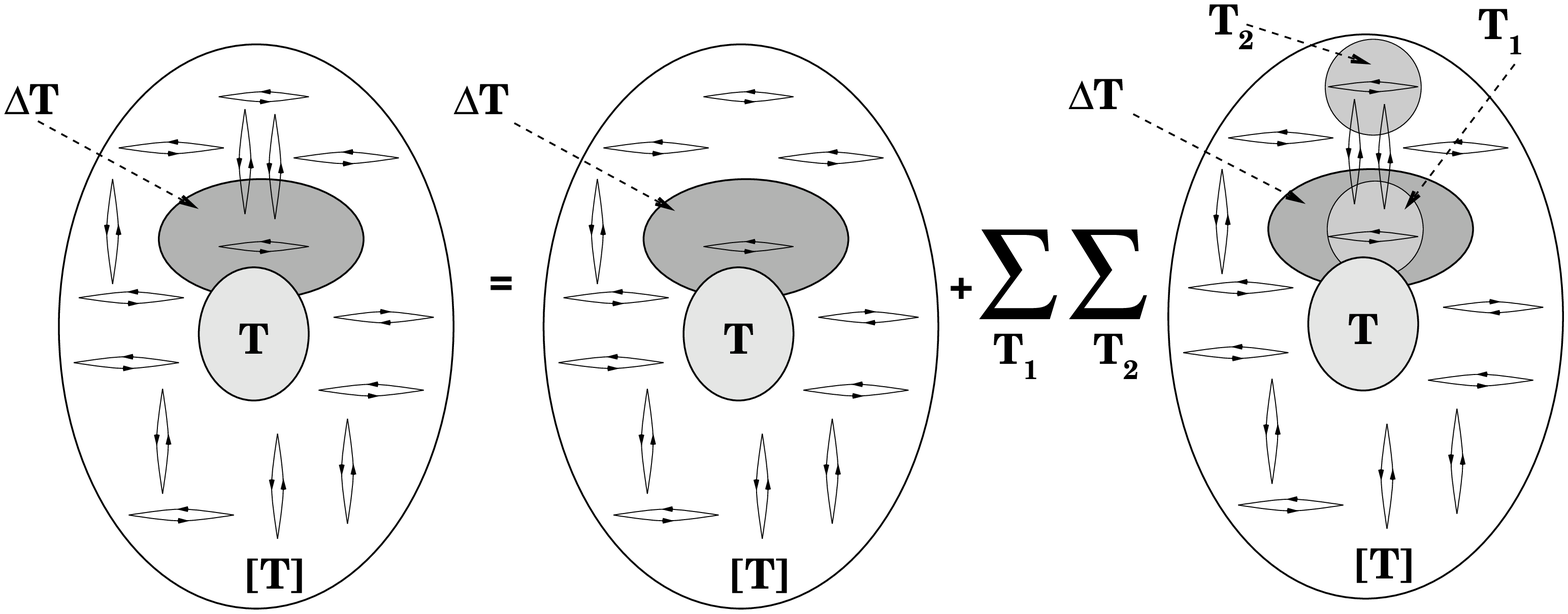}\end{center}

\caption{Schematic illustrating Eq. (\ref{eq:ST-expansion-general}): all
AD's corresponding to the manifold $[T]$ of electronic groups can
be represented as a sum of contributions based on choosing an arbitrary
sub-manifold of groups $\Delta T\subset[T]$. \label{cap:Schematic}}
\end{figure}

The corresponding relationship between the two pre-factors can be
obtained as follows. Consider the sum of all possible AD's which
make up the entire normalisation integral $\widetilde{S}[T]$ for
the subsystem $[T]$. From all of them we separate out those which
are related to the set $[T+\Delta T]$:\begin{equation}
\widetilde{S}[T]=\widetilde{S}(\Delta T)\widetilde{S}[T+\Delta
T]+\sum_{T_{1}\subset\Delta T}\:\sum_{T_{2}\subset\left[T+\Delta
T\right]}\widetilde{D}(T_{1}T_{2})\widetilde{S}(\Delta
T-T_{1})\widetilde{S}[T+\Delta
T+T_{2}]\label{eq:ST-expansion-general}\end{equation} The first
term in the right hand side contains all the AD's that can be
constructed out of the groups of the set $\Delta T$, given by
$\widetilde{S}(\Delta T)$. These should be multiplied by all AD's
constructed out of the rest of the groups, i.e. by all groups from
the manifold $[T+\Delta T]$. The second term contains all the AD's
which are formed by the groups belonging to \emph{both} sets
$\Delta T$ and $[T+\Delta T]$. Indeed, the double sum picks up
subsets $T_{1}$ and $T_{2}$ from the manifolds of groups $\Delta
T$ and $[T+\Delta T]$, respectively. The sum of all possible AD's
constructed using \emph{every} group from $T_{1}$ and $T_{2}$ is
denoted $\widetilde{D}(T_{1}T_{2})$. Note that AD's in
$\widetilde{D}(T_{1}T_{2})$ may contain non-linked diagrams as
well; however, in this case \emph{each} of non-linked AD's must
 contain \emph{all} groups from $T_{1}$ and $T_{2}$. Every AD in
$\widetilde{D}(T_{1}T_{2})$ is multiplied by all possible AD's
formed out of the rest of groups of the two sets:
$\widetilde{S}(\Delta T-T_{1})$ corresponds to the AD's made of
all groups in the set $\Delta T-T_{1}$ (created by removing groups
belonging to $T_{1}$ from the set $\Delta T$), whereas
$\widetilde{S}[T+\Delta T+T_{2}]$ is obtained by all AD's left
after {}``removing'' the set $T_{2}$ from the set $[T+\Delta T]$.
Eq. (\ref{eq:ST-expansion-general}) is schematically illustrated
in Fig. \ref{cap:Schematic}. %

Dividing each term in Eq. (\ref{eq:ST-expansion-general}) by $\widetilde{S}$,
we obtain a recurrence relation for the pre-factors sought for: \begin{equation}
f_{[T]}=\widetilde{S}(\Delta T)f_{[T+\Delta T]}+\sum_{T_{1}\subset\Delta T}\:\sum_{T_{2}\subset\left[T+\Delta T\right]}\widetilde{D}(T_{1}T_{2})\widetilde{S}(\Delta T-T_{1})f_{[T+\Delta T+T_{2}]}\label{eq:fT-expansion-general}\end{equation}

It is seen that $f_{[T]}$ is represented as a linear combination
of pre-factors corresponding to \emph{smaller} systems: if $\aleph_{T}$
is the number of groups in the manifold $T$, then we can write that
$\aleph_{[T]}>\aleph_{[T+\Delta T]}>\aleph_{[T+\Delta T+T_{2}]}$.
Eq. (\ref{eq:fT-expansion-general}) is a generalisation of any of
the Eqs. (\ref{eq:for_S}) - (\ref{eq:for_f123}) written above for
the toy model.

Equations like the one written above allow one to obtain the necessary
series expansion in the general case. To this end, we attach to every
diagram a factor $\sigma^{n}$ with the power $n$ being the order
of the AD, i.e. the number of arrows it contains. The expansion we
obtain will be with respect to $\sigma$ that will be set to unity
at the end of the calculation. Note that this method does not take
into account an additional factor, which is relevant in the actual
calculation, that the contributions of the AD's also depend on the
distances between different groups. In the method we propose an expansion
with respect to overlap integrals will be obtained.

\begin{figure}
\begin{center}\includegraphics[%
  height=4cm]{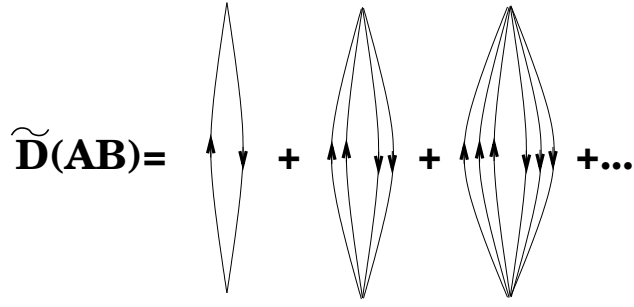}\end{center}

\caption{AD's representation of $\widetilde{D}(AB)$ of Eq.
(\ref{eq:exp_for_S(AB)}).\label{cap:ADs-represent-of-Sab}}
\end{figure}

\begin{figure}
\begin{center}\includegraphics[%
  height=3cm]{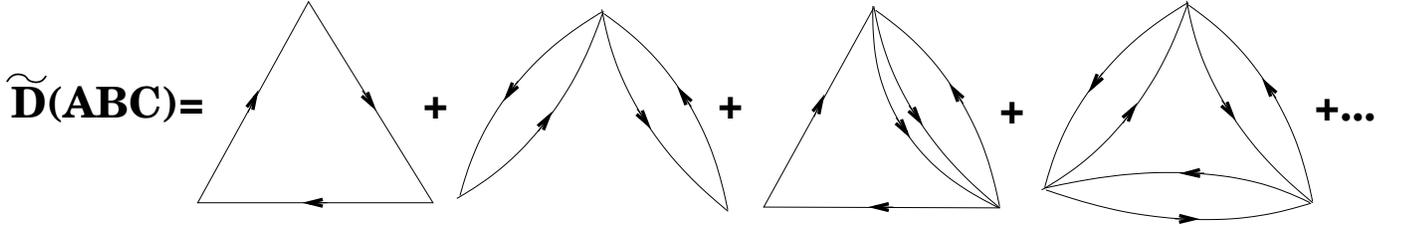}\end{center}

\caption{AD's representation of $\widetilde{D}(ABC)$ of Eq.
(\ref{eq:exp_for_S(ABC)}).\label{cap:ADs-represent-of-Sabc}}
\end{figure}

To illustrate how the method works, we shall obtain a few first
terms in the expansion for $f_{[A]}$, $f_{[A,B]}$ and
$f_{[A,B,C]}$. Noting that $\widetilde{S}(A)=1$, we start,
similarly to section \ref{sub:Series-expansion-for-toy-model},
with $T=\emptyset$ (empty) and $\Delta T=A$ in Eq.
(\ref{eq:fT-expansion-general}):\begin{equation}
1=f_{[A]}+\sum_{B\in[A]}\widetilde{D}(AB)f_{[A,B]}+\sum_{\{
B,C\}\in[A]}\widetilde{D}(ABC)f_{[A,B,C]}+\sum_{\{
B,C,D\}\in[A]}\widetilde{D}(ABCD)f_{[A,B,C,D]}+\ldots\label{eq:general-f0}\end{equation}
where the curly brackets under the summation sign indicate that we
sum over all \emph{sets} of the groups containing two, three, etc,
groups in them irrespective of their order in the sets;
further,\begin{equation}
\widetilde{D}(AB)=D_{2}(A,B)\sigma^{2}+D_{4}(A,B)\sigma^{4}+\ldots\label{eq:exp_for_S(AB)}\end{equation}
\begin{equation}
\widetilde{D}(ABC)=D_{3}(A,B,C)\sigma^{3}+D_{4}(A,B,C)\sigma^{4}+D_{5}(A,B,C)\sigma^{5}+\ldots\label{eq:exp_for_S(ABC)}\end{equation}
contain all AD's composed of groups $\{ A,B\}$ and $\{ A,B,C\}$,
respectively. For convenience, contributions of different orders
with respect to $\sigma$ have been presented separately by the
quantities like $D_{n}(T)$. In particular, $\widetilde{D}(AB)$
contains a sum of bubble AD's which only have even number of
arrows as shown in Fig. \ref{cap:ADs-represent-of-Sab}. %
 The expansion is finite: it runs until the number of electrons in
either of the groups is completely used up. For instance, if a
group $A$ contains only two electrons, there will only be the
first two terms in the expansion. Similarly, $\widetilde{D}(ABC)$
contains a finite sum of all AD's made out of groups $\{ A,B,C\}$
as shown in Fig. \ref{cap:ADs-represent-of-Sabc} (note that both
directions
of arrows, where appropriate, are assumed in this expansion). %
 The expansion starts from the AD containing 3 arrows, the next term
contains 4 arrows, then follows the term with 5 arrows, etc. For
small values of $n$ there might be only few AD's contained in the
given $D_{n}(T)$; however, the number of the AD's may grow quickly
as the order $n$ becomes larger. The only exception is
$\widetilde{D}(AB)$ in which case in any order $n$ (that is always
even) there is exactly one AD in $D_{n}(T)$.

Following the same logic, consider Eq.
(\ref{eq:fT-expansion-general}) for $T=A$ and $\Delta T=B$. This
gives: \begin{equation}
f_{[A]}=f_{[A,B]}+\sum_{C\in[A,B]}\widetilde{D}(BC)f_{[A,B,C]}+\sum_{\{
C,D\}\in[A,B]}\widetilde{D}(BCD)f_{[A,B,C,D]}+\ldots\label{eq:general-fa}\end{equation}
If $T=\{ A,B\}$ and $\Delta T=C$, then we also have:
\begin{equation}
f_{[A,B]}=f_{[A,B,C]}+\sum_{D\in[A,B,C]}\widetilde{D}(CD)f_{[A,B,C,D]}+\sum_{\{
D,E\}\in[A,B,C]}\widetilde{D}(CDE)f_{[A,B,C,D,E]}+\ldots\label{eq:general-fab}\end{equation}
This process can be continued: at every next step one chooses the
set $T+\Delta T$ of the previous step as the set $T$, and one
group from the new $[T]$ is chosen as the set $\Delta T$.

Expanding all the pre-factors in terms of $\sigma$ (see Eq.
(\ref{eq:expansion_for_f})) and comparing terms with the same
powers of $\sigma$, we obtain after setting $\sigma=1$: \[
f_{[A]}=1-\sum_{B\in[A]}D_{2}(A,B)-\sum_{\{
B,C\}\in[A]}D_{3}(A,B,C)+\sum_{B\in[A]}\left[D_{2}(A,B)\right]^{2}\]
\[
+2\sum_{\{
B,C\}\in[A]}D_{2}(A,B)\left[D_{2}(A,C)+D_{2}(B,C)\right]-\sum_{B\in[A]}D_{4}(A,B)-\sum_{\{
B,C\}\in[A]}D_{4}(A,B,C)-\sum_{\{ B,C,D\}\in[A]}D_{4}(A,B,C,D)\]
\[
+\sum_{\{ B,C\}\in[A]}\left\{
D_{3}(A,B,C)\left[4D_{2}(A,B)+D_{2}(B,C)\right]-D_{5}(A,B,C)\right\}
-\sum_{\{ B,C,D,E\}\in[A]}D_{5}(A,B,C,D,E)\]
\begin{equation}
+\sum_{\{ B,C,D\}\in[A]}\left\{
6D_{3}(A,B,C)\left[D_{2}(A,D)+D_{2}(B,D)\right]+3D_{3}(B,C,D)D_{2}(A,D)-D_{5}(A,B,C,D)\right\}
+\ldots\label{eq:final_fA}\end{equation}

The expression for the pre-factor $f_{[A]}$ was obtained up to the
5-th order with respect to overlap and with the terms ordered
appropriately. Expansions for $f_{[A,B]}$ and $f_{[A,B,C]}$ are
obtained up to the fourth order with respect to overlap using the
same equations (\ref{eq:general-f0}), (\ref{eq:general-fa}) and
(\ref{eq:general-fab}). Because in the final expression for the
RDM-1 the pre-factors are multiplied by the linked AD's (with the
open circle) which are also of some order with respect to the
overlap, third and second order expansions are needed for
$f_{[A,B]}$ and $f_{[A,B,C]}$, respectively, if the same order
expression for the RDM-1 is to be derived. Then, one obtains:\[
f_{[A,B]}=1-D_{2}(A,B)-\sum_{C\in[A,B]}\left[D_{2}(A,C)+D_{2}(B,C)\right]-\sum_{C\in[A,B]}D_{3}(A,B,C)\]
\begin{equation}
-\sum_{\{
C,D\}\in[A,B]}\left[D_{3}(A,C,D)+D_{3}(B,C,D)\right]+\ldots\label{eq:final_fAB}\end{equation}
\begin{equation}
f_{[A,B,C]}=1-\left[D_{2}(A,B)+D_{2}(B,C)+D_{2}(A,C)\right]-\sum_{D\in[A,B,C]}\left[D_{2}(A,D)+D_{2}(B,D)+D_{2}(C,D)\right]+\ldots\label{eq:final_fABC}\end{equation}
It is seen that in all orders the first-order term vanishes, i.e.
$a_{1}^{[T]}=0$.

Note that the choice of manifolds $T$ and $\Delta T$ is somewhat
arbitrary. Any choice results in the same expansions. However, the
method outlined above is, in our view, the simplest one which leads
to a systematic derivation of the expansion coefficients for the pre-factors
of increasingly higher orders.

\section{Conclusions}

In this paper we have re-examined the {}``linked-AD theorem'' of
Ref. \cite{linked-AD} formulated for extended (infinite) systems
containing a very large number of electronic groups. For example,
in the case of the electron density, the theorem states that the pre-factors,
$f_{[T]}$, to the linked AD's with an open circle (involving groups
of the set $T$) are all equal to unity. By analysing in detail an
exactly solvable system (1D Hartree-Fock ring with a single electron
in each group and the nearest neighbours non-zero overlap), we find,
however, that this theorem is only \emph{approximately} valid when
the group wavefunctions are \emph{strongly localised}. When this is
not the case, so that the overlap between neighbouring groups is significant,
corrections to the theorem have to be applied and the pre-factors
may strongly deviate from the value of unity.

\begin{figure}
\begin{center}\includegraphics[%
  height=4cm]{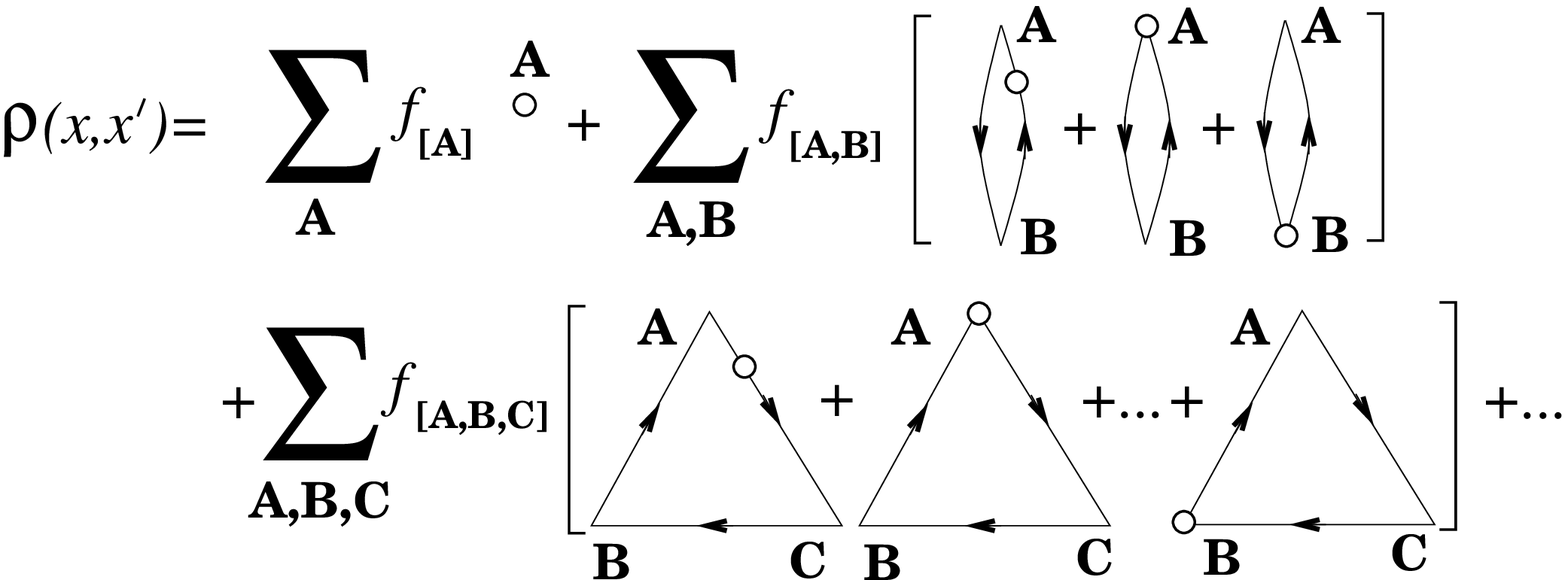}\end{center}

\caption{The first three terms in the AD expansion for the RDM-1. \label{cap:RDM-1}}
\end{figure}

\begin{figure}
\begin{center}\includegraphics[%
  height=4cm]{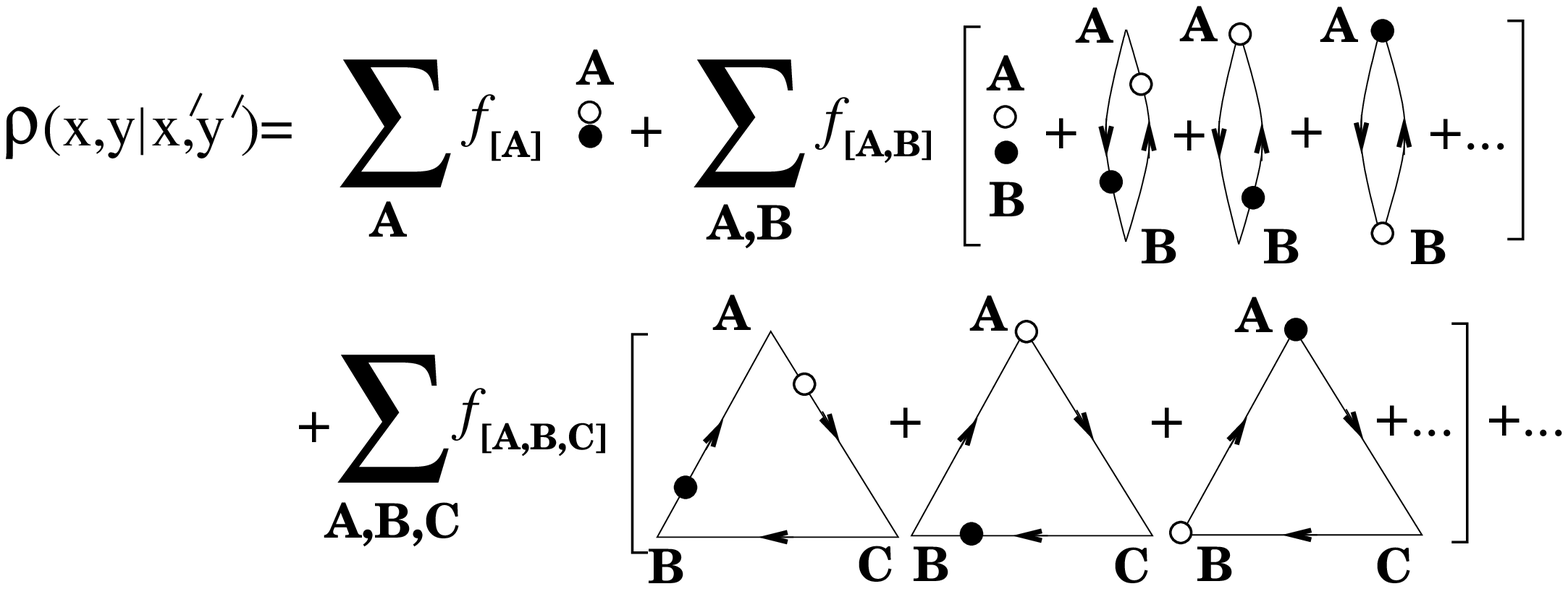}\end{center}

\caption{A few first terms in the AD expansion for the RDM-2.
\label{cap:RDM-2}}
\end{figure}

To obtain such corrections, a general method based on the
expansion of the pre-factors in power series with respect to
overlap between different groups, has been suggested. To
illustrate the method, we have obtained in detail the corrections
for the first three pre-factors in the expression for the RDM-1,
namely: (i) $f_{[A]}$ (associated with the circle AD); (ii)
$f_{[A,B]}$ (a bubble AD with an open circle on an arrow) and
(iii) $f_{[A,B,C]}$ (a triangle AD with an open circle
on an arrow), as shown in Fig. \ref{cap:RDM-1}. %
Other pre-factors can be obtained along the same lines. Note that
expressions for the pre-factors do not depend on the particular
type (topology $q$, see section \ref{sec:The-AD-theory}) of the AD
with an open circle; they only depend on the particular groups
(the set $T$) involved in the diagram. Due to this, \emph{the
same} pre-factors will appear in the expansion of the RDM-2 as
well. Several first terms in the expansion of the latter are shown
in Fig. \ref{cap:RDM-2}. The filled and open black circles
designate the variables $(x,x^{\prime})$ and $(y,y^{\prime})$ of
the RDM-2
$\rho\left(x,y\left|x^{\prime},y^{\prime}\right)\right.$,
respectively (see \cite{linked-AD,my-AD-2} for more details).%

Therefore, an energy expression up to arbitrary order with respect
to overlap can finally be obtained taking into account the described
corrections to the linked-AD theorem. It is different from that presented
previously \cite{linked-AD} only by the appearance of the pre-factors,
so that we do not show it here in detail.

It is relevant to note that an expansion of the electron density with
respect to the overlap in the Hartree-Fock case (with no electron
correlation within groups included) is well known \cite{McWeeny}.
It is obtained from Eq. (\ref{eq:ro-via-inverse-S}) by representing
the overlap matrix $\mathbf{S}$ as $\mathbf{S=1}+\mathbf{\Delta}$
and then expanding $\mathbf{S}^{-1}=\left(\mathbf{1}+\mathbf{\Delta}\right)^{-1}$
with respect to the matrix $\mathbf{\Delta}$; the latter contains
only overlap between different groups and zeros along the diagonal.
The method developed in this paper in section \ref{sub:General-method}
generalises this expansion method to the case when group functions
are linear combinations of Slater determinants, i.e. when intra-group
correlation effects are accounted for in each group. However, it is
also known (see, e.g. \cite{Danyliv-LK-periodic-2004}) that the overlap
expansion of the electron density in the HF case \emph{diverges} if
the overlap between groups is significant (more explicitly, if there
is at least one eigenvalue of $\mathbf{\Delta}$ which is larger than
unity). Therefore, our method should also have some limitations on
its convergence and must be applied with care.


\end{document}